# Tunable Optical Radiation Formed in Crystalline Undulator

**Hayk Gevorgyan**[a,b,*]

[a]A. I. Alikhanian National Science Laboratory (Yerevan Physics Institute), 2 Alikhanian Br. Street, Yerevan, Armenia, 0036
[b]Yerevan State University, Physics Faculty, Theoretical Department, 1 Alex Manoogian, Yerevan, Armenia, 0025

**Abstract**. The problem of optical radiation produced by a bunch of relativistic positrons channeled to a crystalline undulator (CU) is considered. Taking into account the polarization of the crystal medium, an energy threshold takes place to form the radiation. If the energy of the bunch is above the threshold, the radiation is created in a limited frequency range. Both soft and hard boundary photons exist and are emitted at zero angle due to the polarization of the medium. In the case where the ratio between bunch's and threshold energies is much greater than unity, the energy of the boundary soft photons depends only on the characteristic parameters (spatial period, amplitude) of the crystalline sinusoidal undulator and can be tuned to the optical range. It is shown that it is possible to obtain a directed optical beam in the "water window" range. Such photon beams have important applications in medicine and biological research.

## 1 Introduction

If CU has the sinusoidal form, then the trajectory of the oscillating positrons, channeled between the crystal planes, on average has the sinusoidal form too. In addition to the channeling radiation, is formed CU radiation. The idea of using an oscillating relativistic bunch of charged particles for the optical radiation formation belongs to Ginsburg [1]. The problem of the X-ray undulator radiation was considered by Korkhmazyan [2]. The gamma radiation of the particles channeled in crystal was investigated by Kumakhov [3]. The channeling phenomena was explained by Lindhard [4]. The characteristics of undulator radiation in the dispersing medium are revealed [5]. In the monograph the different types of the spontaneous electromagnetic radiation was considered [6]. The generation of radiation using CU was proposed in [7]. The spectral distribution of radiation, formed in CU, without [8] and with [9] taking into account the medium polarization was investigated. The modulated bunch coherent radiation at a resonant frequency, formed in the dispersing medium of CU, was obtained [10]. In this work is investigated the possibility of obtaining a sufficiently intense monochromatic directional beam of photons with the energy 0.28÷0.54 KeV or with the wavelength 4.4÷2.3 nm. In the water and in the carbon the absorption length of these photons is more and less, accordingly, than for the other photons.

## 2  The Conservation of Planar Channeling of Positrons in CU

The relativistic positrons, falling parallel to the planes of crystal, interact with many atoms of crystal simultaneously. Therefore, between the crystal planes they move in the average potential field of these atoms. For this field the harmonic potential is a good approximation [6]. The positrons are oscillating according to the harmonic law. The positrons are oscillating according to the harmonic law, with the frequency depending on their energy. This phenomena of channeling is conserved at falling angles smaller than the Lindhard critical angle $\Theta_L = \sqrt{2\nu/\gamma}$, where $\nu$ and $\gamma$ are the potential well depth and the positron energy in units of the rest energy of a positron $mc^2$, accordingly. At large falling angles, the transverse kinetic energy becomes more than the potential well depth and the movement becomes infinite.

Let the longitudinal section of CU planes has the sinusoidal form. Then the channeling phenomena will take place for the all positrons with the normal vector of falling, if the maximal angle of the crystal bending is smaller than the angle Lindhard $\Theta_{max} < \Theta_L$. CU parameter is defined with the following way $q_{cu} = \Theta_{max}\gamma$. The definition of a channeling parameter thus $q_{ch} = \Theta_L \gamma$ is leading to a sufficient condition for the existence of CU $q_{cu} < q_{ch}$ (the channeling conservation condition in CU). For the sinusoidal CU we have $\Theta_{max} = 2\pi A/l$, where $A$ and $l$ are the amplitude and the space period of CU accordingly, therefore, the limiting condition for CU parameter is $A/l < \sqrt{(2\nu/\gamma)}/(2\pi)$.

## 3 The Photon Number Spectral Distribution, Formed in CU

The radiation wave vector depends on the electric permittivity of medium. Since the radiation generation occurs in the dispersing medium of CU, then permittivity is given by the following formula:

$$\varepsilon(\omega) = 1 - \left(\frac{\omega_p}{\omega}\right)^2, \qquad \omega \gg \omega_p, \qquad (1)$$

where $\omega$ is the radiation frequency, and $\omega_p$ is the medium plasma frequency. Ignoring that a positron velocity $V$ is changed at the radiation process, and supposing that it moves with the average longitudinal constant velocity $V_\parallel$ in CU (dipole approximation), for the number frequency-angular distribution of radiated photons turns out the following formula [5]

$$\frac{d^3 N_{ph}}{d\omega d(\cos\theta) d\varphi} = \frac{\alpha L^2 \sqrt{\varepsilon}}{16\pi^2 V^2} \sum_{k=1}^{\infty} \left( \beta_\parallel^2 \sin^2\theta - \frac{k}{a} \beta_\parallel \beta_\perp \sin 2\theta \sin\varphi + \frac{k^2}{a^2} \beta_\perp^2 (1 - \sin^2\theta \sin^2\varphi) \right) \mathfrak{J}_k^2(a) F(z_k),$$

$$a = \frac{\omega}{\Omega} \beta_\perp \sqrt{\varepsilon} \sin\theta \sin\varphi, \qquad F(z_k) = \frac{\sin^2 z_k}{z_k^2}, \qquad z_k = \pi n \left( k - \frac{\omega}{\Omega}(1 - \beta_\parallel \sqrt{\varepsilon} \cos\theta) \right), \qquad (2)$$

where $\alpha = 1/137$ is the fine structure constant, $L = nl$ is CU length, $n$ is CU period number (oscillation number of a positron), $\Omega$ is CU frequency (oscillation frequency of a positron), $\beta_\parallel = \frac{V_\parallel}{c}, \beta_\perp = \frac{V_\perp}{c} \approx \Theta_{max}$, $c$ is the light velocity in vacuum, $V_\perp$ is the maximum value of the positron transverse velocity, $\theta$ and $\varphi$ are the polar angle and the azimuthal angle of radiation accordingly, $k$ is the harmonic number, $\mathfrak{J}_k$ is the Bessel function of the first kind of order $k$.

In the laboratory frame due to the Doppler effect, the radiation frequency of a relativistic positron is much more than CU frequency ($\omega \gg \Omega$). Therefore, the energy & impulse conservation law ($z_k = 0$) is satisfied at the small angles of radiation ($\theta \ll 1$). The objective of this work is the obtaining directional radiation (at the zero angle). At a zero angle is formed the radiation of first harmonic only.

The photon number frequency-angular distribution for the first harmonic radiation have the following form

$$\frac{d^3 N_{ph}}{d\omega d(\theta^2) d\varphi} = \frac{\alpha \omega}{2} \left(\frac{n q_{cu}}{2\omega \gamma}\right)^2 \left(\cos^2 \varphi + \left(1 - \theta^2 \frac{\omega}{\Omega}\right)^2 \sin^2 \varphi\right) \frac{\sin^2 z}{z^2}, \quad (3)$$

$$z(\omega, \theta) = \frac{\pi n \omega}{2\Omega \gamma^2} \left(\left(1 - \frac{\omega_{min}}{\omega}\right)\left(\frac{\omega_{max}}{\omega} - 1\right) - \theta^2 \gamma^2\right),$$

$$\omega_{\substack{min \\ max}} = \frac{\Omega \gamma^2}{Q} \left(1 \mp \sqrt{1 - \left(\frac{\gamma_{th}}{\gamma}\right)^2}\right),$$

$$Q = 1 + \frac{q_{cu}^2}{2},$$

$$\gamma_{th} = \sqrt{Q} \frac{\omega_p}{\Omega},$$

where $\gamma_{th}$ is the energy threshold for the radiation generation in the dispersing medium.

Let us consider the case of $\gamma \gg \gamma_{th}$. Then, with accuracy up to small order $(\gamma_{th}/\gamma)^2$, we have the following

$$\omega_{min} = \frac{\omega_p^2}{2\Omega} = \frac{\Omega}{2Q} \gamma_{th}^2, \qquad \omega_{max} = \frac{\Omega}{2Q} \gamma^2, \qquad \omega_{max} \gg \omega_{min}. \quad (4)$$

The photon number frequency distribution at the zero angle is the following

$$\frac{dN_{ph}}{d\omega} = \frac{\pi \alpha}{Q} (q_{cu} n)^2 \frac{\omega \omega_{min}}{\omega_p^2 \omega_{max}} \frac{\sin^2 z(\omega, 0)}{z(\omega, 0)}, \quad (5)$$

$$z(\omega, 0) = \frac{\pi n}{Q} \left(1 - \frac{\omega_{min}}{\omega}\right)\left(1 - \frac{\omega}{\omega_{max}}\right).$$

The relative linewidth of the diffractive sine is equal to $\Delta\omega/\omega_{min} = \Delta\omega/\omega_{max} = Q/n$.

The soft radiated photon number with the energy $\hbar\omega_{min}$ and with the relative energy spread $Q/n$ is equal to

$$N_{ph}(\omega_{min}) = \pi \alpha n q_{cu}^2 \frac{\omega_{min}^3}{\omega_p^2 \omega_{max}}. \quad (6)$$

## 4 The Generation of Directed Beam of Optical Photons

For generation of optical photons with the frequency $\omega_{min}$ ($\lambda_{max}$) is necessary to use CU with the following space period $l$ and amplitude $A$:

$$l = \frac{2\lambda_p^2}{\lambda_{max}}, \qquad A < \frac{l}{\pi}\sqrt{\frac{v}{2\gamma}}. \tag{7}$$

This choice of $A$ corresponds to the channeling conservation condition in CU. Since intensity of photons (the photon number) is proportional to $q_{cu}^2$, then for generation of the intensive radiation is necessary to provide the following value $q_{cu} \approx 1$ at least.

Let the positron bunch has the following parameters (we believe that it is possible to obtain a positron bunch with the same parameters as the LCLS electron bunch): the positron number in bunch is $N_b = 1.56 \cdot 10^9$, the positron energy (bunch energy) is $E = 13.6$ GeV ($\gamma = 2.66 \cdot 10^4$), the transverse section is $S = 1.18 \cdot 10^6$ cm² ($\sigma_r = 6.12 \cdot 10^{-4}$ cm).

CU represents the periodically bent monocrystal of diamond with the space period $l = 9 \cdot 10^{-5}$ cm and with the amplitude $A = 0.085$ Å ($q_{cu} = 1.6$, $Q = 2.28$), the crystallographic planes (1 1 0) of which have the curved sinusoidal form. The energy of the plasma oscillations of the diamond medium is equal to $\hbar\omega_p = 38$ eV ($\lambda_p = 3.26 \cdot 10^{-6}$ cm). For the diamond and (1 1 0) planes, the depth of potential well is $U_0 = 24.9$ eV ($v = \frac{U_0}{mc^2} = 4.87 \cdot 10^{-5}$). The condition $q_{cu} < q_{ch}$ was satisfied since $q_{ch} = \sqrt{2v\gamma} = 1.62$. Then, on the interact length $L = 1$ cm ($n = 1.1 \cdot 10^4$) will be generated the beam consisting of $N_{tot} = N_b \cdot N_{ph} = 10^8$ photons with the energy $\hbar\omega_{min} = 0.5$ KeV ($\lambda_{max} = 2.35$ nm) and with the relative energy spread of the order $Q/n \sim 10^{-4}$. In the result we will obtain the directional beam at the zero angle with the surface density of the order $N_{tot}/S \sim 10^{14}$ cm⁻².

## 5    Conclusion

It was shown that CU with the certain parameters can serve as the compact radiator for the generation of optical photons, using the relativistic positron bunch. The considering of a certain example was shown the possibility of generation of the monochromatic, intense and directed photon beam with energy from the "water window" region. Such beams of photons are necessary for the biological research and for the practical application in medicine.

*References*

1. V. L. Ginzburg, "On radiation of micro-radio-waves and their absorption in air," *Izvestiya Akad. Nauk SSSR, Ser. Fiz.* **11**, 165-182 (1947).

2. N. A. Korhmazyan, "Radiation of Fast Charged Particles in Transverse Sinusoidal Electric Fields (in Russian)," *Izv. Akad. Nauk Arm. SSR*, *Ser. Fiz.* **5**, 287-8 & 418 (1970).

3. M. A. Kumakhov, "On the theory of electromagnetic radiation of charged particles in crystal," *Phys. Lett.* **57A**, 17-18 (1976).

4. J. Lindhard, "The influence of crystal lattice on the motion of fast charged particles," *Physcis Uspekhi* **99**(2) **N2**, 249-296 (1969).

5. L. A. Gevorgyan, N. A. Korkhmazyan, "Undulator radiation in dispersive media," *JETP* **76**, 1226-1235 (1979).

6. V. A. Bazylev, N. K. Zhivago, "Radiation from fast particles in external fields," *Nauka*, *Moscow*, (1987).

7. V. V. Kaplin, S. V. Plotnikov, S. A. Vorobev, "Radiation by charged particles in deformed crystal," *Zh. Tekh. Fiz.* **50**, 1079-1081 (1980)

8. A. A. Korol, A. V. Solov'yov, W. Greiner, "Coherent radiation of an ultrarelativistic charged particle channeled in a periodically bent crystal,"*J. Phys. G: Nucl. Part. Phys.* **24** (5), L45-L53, (1998)

9. R. O. Avakian, L. A. Gevorgian, K. A. Ispirian, R. K. Ispirian, "Spontaneous and stimulated radiation of particles in crystalline and nanotube undulators," *NIM B* **173**, 112-120 (2001)

10. L. A. Gevorgian, K. K. Ispirian, A. H. Shamamian, "Crystalline undulator radiation of microbunched beams taking into account the medium polarization," *NIM B* **309**, 63-66 (2013)